\documentclass[pra,twocolumn,aps,amssymb,footinbib,showpacs,]{revtex4}
\usepackage{amssymb}

\usepackage{graphicx}
\usepackage{amsmath}
\usepackage{times}

\begin{document}

\title{Hanbury Brown-Twiss Interferometry for Fractional and Integer Mott Phases}

\author{ Ana Maria Rey$^{1}$, Indubala I. Satija$^{2,3}$ and Charles W. Clark$^{2}$} \affiliation{$^{1}$ Institute for
Theoretical Atomic, Molecular and Optical Physics, Cambridge, MA,
02138,USA} \affiliation{ $^{2}$ National Institute of Standards and
Technology, Gaithersburg MD, 20899, USA} \affiliation{$^{3}$ Dept.
of Phys., George Mason U., Fairfax, VA, 22030,USA}

\date{\today}
\begin{abstract}

Hanbury-Brown-Twiss interferometry (HBTI) is used to study integer
and fractionally filled Mott Insulator (MI) phases in  period-$2$
optical superlattices. In contrast to the quasimomentum
distribution, this second order interferometry pattern exhibits high
contrast fringes in the {\it insulating phases}. Our detailed study
of HBTI suggests that this interference pattern signals the various
superfluid-insulator transitions and  therefore can be used as a
practical method to determine the phase diagram of the system. We
find that in the presence of a confining potential   the insulating
phases become robust as they exist  for a finite range of atom
numbers. Furthermore, we  show that in the trapped case the HBTI
interferogram signals the formation of the MI domains and probes the
shell structure of the system.

\end{abstract}
\pacs{71.30.+h, 03.75.Lm, 42.50.Lc} \maketitle

\section{Introduction}

Cold atoms in optical lattices are becoming ideal many-body systems
to attain laboratory demonstrations of  model  quantum Hamiltonians
due in part to the dynamical  experimental control of the various
parameters at a level unavailable in more traditional condensed
matter systems. Recent experiments with bosonic atoms have been able
to simulate effective one dimensional systems
\cite{Tolra,Weiss,Paredes,Moritz,Fertig}, to enter the strongly
correlated Tonks Girardeau (TG) regime \cite{Weiss,Paredes}
predicted many years ago \cite{GR} and to realize the  superfluid to
Mott insulator(MI) \cite{Fisher} transition by tuning the lattice
parameters \cite{Greiner}. In fact, one of the most active frontier
in cold atom systems is to explore the possibility of creating  new
quantum phases\cite{newphases}. The capability to physically create
them demands sophisticated diagnostic tools for their
characterization and actual observation in the laboratory. In this
paper, we use second order interference techniques, namely the
Hanbury Brown and Twiss interferometry\cite{HBT} to describe the
rich phase diagram of interacting bosonic atoms in the presence of
two competing lattices.

Multiple well super-lattices have  been experimentally realized by
superimposing  two independent  optical lattices with different
periodicities\cite{Drese, Peil}. Recent theoretical studies of the
many-body Bose-Hubbard (BH) Hamiltonian in the presence of an
additional superlattice \cite{Roth,Buonsante1, Buonsante2,
Buonsante3, Buonsante4, Rigol} have centered mostly on the detailed
phase diagrams of the system. The focus of this paper is to
characterize the different phases of the system by analyzing the
four and two point correlations that can be extracted from the time
of flight images. Our study is confined to period-$2$ superlattices,
the simplest example of a system exhibiting almost all the key
aspects of a more general period-$m$ lattices.

The HBTI technique  measures second order correlations from the
shot-noise density fluctuations in the absorption images of
expanding atomic gas clouds. This technique has been shown
\cite{Altman, Foelling, Greiner2} to be  particularly suited for
probing many-body states of cold atomic systems, as it  provides
complementary information to the first-order correlations inferred
from the average density distribution in the absorption images. For
example, cold bosonic atoms in the MI regime do not exhibit first
order interference pattern but they do have sharp second-order Bragg
peaks which reflect the spatial periodicity of the lattice. The
interference peaks are the manifestation of the enhanced probability
for simultaneous detection of two bosons (bunching) due to the
Bose-Einstein statistics.

The phase diagram of interacting bosons in the presence of 1D
superlattices is landscaped by various quantum  phases. Transitions
to the different  phases can be driven either by  changing the ratio
between the three sets of energy scales in the system: the
interaction energy $U$, the tunneling rate $J$ and the the lattice
modulation $ \lambda$ or by varying the filling factor $\nu$. The
 lattice modulation   induces additional fractional MI phases that occur at
filling factors commensurate with the periodicity of the combined
superlattice. These phases can be  understood in the hard core boson
(HCB) limit where the strongly interacting bosons can be mapped to
noninteracting fermions. The superlattice fragments the single
particle spectrum, inducing band gaps. The filling of a sub-band at
a critical filling factor in the fermionic system results in a band
insulator state that corresponds to a fractionally filled MI state
of the bosonic atoms. The fractional MI phases survive in the soft
core boson limit beyond a critical value of the interaction energy
$U$.

Besides the fractional Mott phases, the interplay between the
interaction energy  and the superlattice potential can also lead to
a new type of integer-filled Mott phases which cannot be understood
in any special limit such as the HCB limit or the pure periodic case
as they appear at finite $U$ values and for finite strengths of the
superlattice potential. These insulating phases are characterized by
a modulated density profile and hence will be referred as {\it
staggered Mott} phases.

  Here we demonstrate that HBTI provides
definite means to monitor the various superfluid to MI transitions
and we use it as a  method to obtain the phase diagram. Firstly, for
a fixed $U$, $J$ and $\lambda$  we vary $\nu$, and find that the
onset to the transition is accompanied by a change in the sign of
the superlattice induced Bragg peaks as the filling factor is
increased beyond the critical ones. Secondly, for a fixed filling
factor $\nu$ and fixed $\lambda/J$, we change the ratio $U/J$ across
the transitions and show that they are signaled by a sharp maximum
in the intensities of the Bragg interference peaks.  We show that
the phase diagram obtained from HBTI is in qualitative agreement
with the recently reported phase diagram calculated from Monte-Carlo
simulations \cite{Buonsante1,Buonsante2,Buonsante3,Buonsante4,Rigol}
and with a mean field phase diagram that we derive analytically.

Another important aspect of our study is to investigate the effects
of a harmonic trap on the various phases. In contrast to the
translationally invariant system where the transition to insulating
phases occurs only at  critical filling factors,  the parabolic
confinement allows for the formation of  fractional and integer
filled MI domains in a finite window of fillings. We  show that
harmonically confined superlattices  in the HCB limit develop a
shell structure \cite{Jaksch} analogous to the one observed in the
absence of any lattice modulation  at moderate values of  the
interaction. The HBTI interferogram provides clear signature of the
formation of the different Mott domains.

The paper is organized as follows. In Sec.~\ref{for}  we define
noise correlations and the model Hamiltonian used in our study. In
Sec.~\ref{hcbs}, we calculate two and four-point correlations in the
HCB case. In Sec.\ref{scb}, we consider the BH system for finite
values of interaction. We first analytically calculate the phase
diagram by using mean field theory ( whose details are described in
Appendix A) and compare it with the phase diagram obtained from
noise correlations. These higher order correlations are calculated
by numerical diagonalization of the  BH Hamiltonian. In
Sec.\ref{trap}, we discuss the effects of the harmonic trap in the
HCB limit.  In Sec.\ref{conc} we state our conclusions.

\section{Noise Correlations}
\label{for}

In a typical experiment, atoms are released by turning off the
external potentials. The atomic cloud expands, and is photographed
after it enters the ballistic regime. Assuming that the atoms are
noninteracting from the time of release, properties of the initial
state can be inferred from the spatial images
\cite{Altman,Roth,reynoise}: the column density distribution image
reflects the initial quasimomentum distribution, $n(Q)$, and the
density fluctuations, namely the {\it noise correlations}, reflect
the quasimomentum fluctuations,  $\Delta(Q,Q')$,

\begin{eqnarray}
\hat{n}(Q)&=&\frac{1}{L}\sum_{j,k} e^{i Q a(j-k)
}\hat{a}_j^\dagger \hat{a}_k, \label{qumo}\\
 \Delta(Q,Q')&\equiv&
\langle\hat{n}(Q) \hat{n}(Q')\rangle
-\langle\hat{n}(Q)\rangle\langle\hat{n}(Q')\rangle. \label{nnoise}
\end{eqnarray}In Eq. (\ref{nnoise}) we have assumed that  both $Q,Q'$ lie inside the first Brillouin
zone.  Here $L$  is the number of lattice sites and $a$ the lattice
constant. In this paper, for simplicity, we focus on the quantity $
\Delta(Q,0)\equiv \Delta(Q)$. In our discussion below,  $N$ denotes
the number of particles in the system and $\nu=N/L$ is the filling
factor. In this paper, we will confine ourselves to the case with
$\nu \le 1$.

The BH Hamiltonian describes bosons in optical lattices when the
lattice is loaded in such a way that only the lowest vibrational
level of each lattice site is occupied and tunneling occurs only
between nearest-neighbor sites \cite{Jaksch}. The 1D  BH Hamiltonian
in the presence of a period 2-superlattice is given by

\begin{eqnarray}
\hat{H}&=& -J\sum_{\langle i,j\rangle
}\hat{a}_i^{\dagger}\hat{a}_{j}+\frac{U}{2}\sum_{j}\hat{n}_j(\hat{n}_j-1)
\notag \\&&+ \sum_{j} 2 \lambda \cos(\pi j) \hat{n}_j + \sum_{j}
\Omega j^2 \hat{n}_j.
\end{eqnarray}
\noindent Here $\hat{a}_j$ is the bosonic annihilation operator of a
particle at site $j$, $\hat{n}_j=\hat{a}_j^{\dagger}\hat{a}_{j}$,
and the sum $\langle i,j\rangle$ is over nearest neighbors. The
hopping parameter $J$, and the on-site interaction energy  $U$ are
functions of the lattice depth. The cosine term describes the onsite
potential  generated by the additional lattice with twice the
periodicity of the  main lattice. Here the main lattice creates the
tight-binding system and the second lattice is assumed to be  a weak
perturbation. In this case atoms in the super-lattice have a hopping
parameter which is independent of the lattice position.  The
parameter $\lambda$ is almost proportional to the depth (in recoil
units  of the main lattice) of the additional lattice (see Ref.
\cite{Drese} for details). The last term takes into account the
parabolic potential with $\Omega$ proportional to the parabolic
trapping frequency.

\section{Hard Core limit}
\label{hcbs}

In the strongly correlated  regime, when $U \to \infty$,  the BH
Hamiltonian  can be replaced by the HCB Hamiltonian \cite{Sachdev},
\begin{eqnarray}
\hat{H}^{(HCB)}&=&-J\sum_j(\hat{b}_j^{\dagger}\hat{b}_{j+1}
+\hat{b}_{j+1}^{\dagger}\hat{b}_{j})+\notag \\&& \sum_{j} 2 \lambda
\cos(\pi j) \hat{n}_j + \sum_{j} \Omega j^2 \hat{n}_j. \label{HCB}
\end{eqnarray}
Here $\hat{b}_j$ is the annihilation operator at the lattice  site
$j$ which satisfies $[\hat{b}_i,\hat{b}_j^{\dagger}]=\delta_{ij}$,
and the on-site conditions
${\hat{b}_j}^2={\hat{b}_j^{\dagger}}{}^2=0$, which suppress multiple
occupancy of lattice sites.

HCB  operators can be linked to  spin-$1/2$ operators by means of
the Holstein-Primakoff transformation \cite{HP}, which maps bosonic
operators into spin operators. The  Holstein-Primakoff
transformation maps the HCB Hamiltonian  into the XY spin- $1/2$
Hamiltonian. The latter can in turn  be mapped onto  a spinless
fermion  Hamiltonian  by  means of the Jordan-Wigner transformation
\cite{Sachdev, LM}.

The Bose-Fermi correspondence can be used to calculate various
many-body observables of the strongly interacting bosonic system in
terms of  the ideal fermionic two-point functions which  can be
written as  $ g_{lm}=\sum_{k=0}^{N-1} \psi_l^{*(k)} \psi_m^{(k)} $.
Here $\psi_j^{(k)}$  are the amplitudes of the single atom
eigenfunctions at site $j$ and $E_{(k)}$  are the single atom
eigenenergies:
\begin{equation}
-J(\psi_{j+1}^{(k)}+\psi_{j-1}^{(k)})+ 2 \lambda \cos(\pi j )
\psi_{j}^{(k)}+ \Omega j^2 \psi_{j}^{(k)} =E_{(k)}
\psi_{j}^{(k)}.\label{sinpa}
\end{equation}

The single-particle spectrum of period-$2$ superlattice can be
obtained by decimating every other site of the tight binding
Eq.~(\ref{sinpa}). The renormalized system at even or odd sites is
described by
\begin{equation}
J^2(\psi_{j+2}^{(k)}+\psi_{j-2}^{(k)}) +(2J^2+4\lambda^2)
\psi_{j}^{(k)} =E_{(k)}^2 \psi_{j}^{(k)}.\label{sinpa1}
\end{equation}  Assuming periodic boundary conditions, the  eigenenergies  are then given by
\begin{equation}
E_{(k)} =\pm 2 \sqrt{ J^2\cos\left(\frac{4\pi
k}{L}\right)+\lambda^2},
\end{equation} where $k=0,1,\dots, L/2$. The effect of the additional lattice potential is to
split the band into two different sub-bands, each with band-width $2
\sqrt{J^2 +\lambda^2}-2\lambda$, separated by an energy gap of $ 4
\lambda$.

Local observables such as the density distribution and energies are
identical for the HCB and non-interacting fermionic  systems. For
example, the HCB ground state energy corresponds to the sum of the
first $N$ single-particle eigenstates. On the other hand,  fermions
and HCBs posses differ  non-local correlation functions. A general
formulation to calculate HCB two-point correlations has been
developed by Lieb and Mattis \cite{LM}. In our earlier studies
\cite{reynoise}, we have generalized  Lieb and Mattis formalism and
have obtained explicit formulas, involving multiple T\"{o}plitz-like
determinants, to compute the four-point correlation functions
required to calculate noise correlations in HCBs. The evaluation of
these determinants, whose order scales with the size of the system,
is in general  complicated and therefore an analytical treatment is
 difficult.  Below we will describe the results obtained by
numerical computation of the explicit formulas discussed in our
earlier study. For information about the exact formulas and various
other relevant details we refer the readers to our earlier paper
\cite{reynoise}.

\begin{figure}[htbp]
\begin{center}
\includegraphics[width=3.5in]{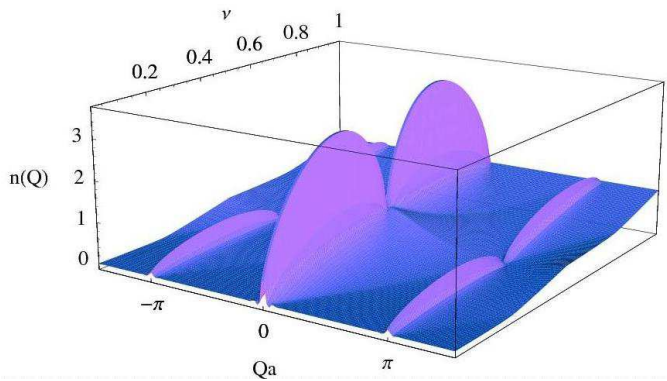} \leavevmode
\end{center}
\caption{Quasimomentum distribution as a function of the filling
factor. The calculations were done for a system with $\lambda/J=1$,
$L=80$ and boxed-like boundary conditions} \label{fig1}
\end{figure}

\begin{figure}[htbp]
\begin{center}
\includegraphics[width=3.5in]{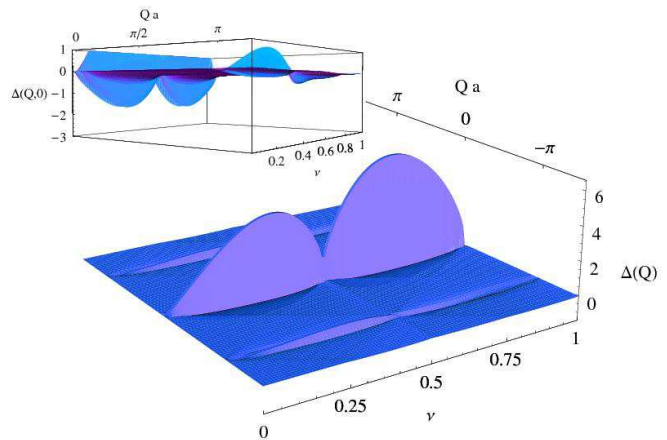} \leavevmode
\end{center}
\caption{Noise interference pattern as a function of the filling
factor for a system with $\lambda/J=1$ , $L=80$ and boxed-like
boundary conditions. In the inset we truncated the central peak to
make the negative background and the peak to dip transition more
visible  } \label{fig2}
\end{figure}

Figs.~\ref{fig1} and \ref{fig2} show the quasimomentum distribution
and the noise correlations for all  filling factors, $ 0 \le \nu \le
1$. The quasimomentum distribution exhibits interference peaks at
the reciprocal lattice vectors of the combined superlattice: a large
peak at $Q=0$ and somewhat weaker peaks at $Qa=\pm \pi$, induced by
the period two modulation. The peaks are a manifestation of  the
quasi-long-range coherence of the system. At $\nu=1/2$ and $1$ the
first order interference pattern flattens out, signaling the
insulating character of the system at these critical fillings.

The peaks at $Qa=0$ and $\pm \pi$ also exist in the noise
interference pattern (Fig.2),  where they are narrower and are
accompanied by adjacent satellite dips, immersed in a negative
background. These satellite  dips are the signatures of the long
range coherence in the second order pattern  as they disappear in
the insulating phases (see detailed discussion in
Ref.\cite{reynoise}). The satellite dips are clearly seen in the
inset.  In contrast to the quasimomentum distribution, the peaks at
reciprocal lattice vectors  continue to exist at the critical
filling factors in the HBTI pattern. However, their intensity is
strongly reduced compared to the intensity at other filling factors.

\begin{figure}[htbp]
\begin{center}
\includegraphics[width=3 in]{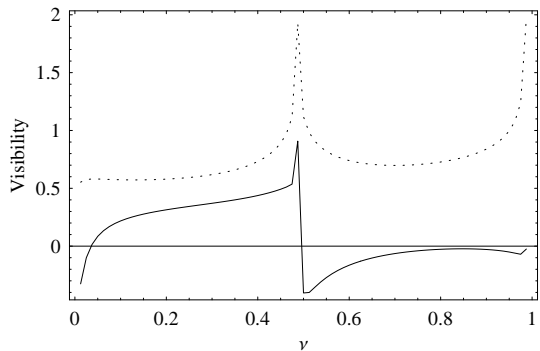} \leavevmode
\end{center}
\caption{Visibility (see text for definition) of the $Q=0$ (dotted
line) and  $Qa=\pi$ (solid lines) peaks for a HCB  system with
$\lambda/J=1$ and $L=80$. } \label{figv}
\end{figure}

To highlight the signatures of  the insulating phases in the noise
correlations,  we plot in Fig.~\ref{figv} the visibility of the
central and superlattice induced peaks. We define the visibility as
the intensity of the second order Bragg fringes (noise-correlations)
normalized with respect to the quasimomentum distribution:
\begin{equation}
{\mathcal{V}}(Q) \equiv \frac{\Delta(Q,0)}{n(Q) n(0)} \label{vis}
\end{equation}
The visibility ${\mathcal{V}(Q)}$ is a relevant experimental
quantity as  the normalization procedure filters some of the
technical noise introduced during  the measurements.

In the visibility  figure one observes the development of  very
sharp peaks at the critical fillings. In addition, an interesting
definitive signal of the fractional insulating phase is a change in
the sign of the intensity of the superlattice induced peaks: as the
filling factor is increased beyond half filling, the peak at
$Qa=\pm\pi$ becomes a dip.  This effect, which we will refer to as
{\it peak to dip transition},  has its origin in the occupation of
the second band with few atoms, once the first band is fully
occupied. In fact, it can be interpreted as a manifestation of the
fermionization of HCBs. The peak to dip transition is a generic
feature of  strongly correlated bosons confined in  a superlattice.
It is seen for all values of $\lambda$ and, as we discuss later,
this signature also accompanies the onset to fractional Mott state
for finite $U$, i.e. in the soft core boson case.

A further insight for this peak to dip transition can be gained by
looking at the noise correlations as $\lambda \rightarrow \infty$ .
In this limit, the correlations can be calculated analytically for
$\nu=1$, $\nu=1/2$  and also for the case of one extra atom beyond
half filling ($N=L/2+1$). At unit filling the ground state  in the
HCB limit corresponds to a unit filled Fock state. In the
half-filled case, to a good approximation, the ground state can be
assumed to be a $1/2$-filled Fock state $|\Psi_g\rangle=|1010\dots
10\rangle$. This is due to the reduced number fluctuations exhibited
in this limit, as the tunneling between every two sites scales as
$J^2/(4\lambda)$. At filling factor $N=L/2+1$ the ground state can
be approximated by a $1/2$-filled Fock state with an extra
delocalized particle at the different unoccupied wells,
$|\Psi_g\rangle=1/\sqrt{L}\sum_{i=1}^N |i\rangle$, with $
|i\rangle=|10\dots11\dots 10\rangle$. Using these ansatzs, it can be
shown that

\begin{eqnarray}
\Delta(Q)_{N=L/2} &=&-\frac{1}{L}+
\frac{1}{4}(3\delta_{Q0}+\delta_{Q\pi}), \label{eq1}\\
\Delta(Q)_{N=L/2+1} &=&-\frac{1}{L}+
\frac{1}{2}(3\delta_{Q0}-\delta_{Q\pi}),\\
 \Delta(Q)_{N=L}
&=&-\frac{2}{L}+2\delta_{Q0}. \label{eq3}
\end{eqnarray}
As shown by these equations, just beyond half-filling, the $Q=0$ and
$Qa=\pi$ fringes have opposite sign. The sign  difference can be
understood by calculating $\Delta(Q)$  for   a single particle
($N=1$) or a single hole ($N=L-1$). In this simple case, noise
correlations can be calculated explicitly for all values of
$\lambda$:
\begin{equation}
\Delta(Q,0)_{N=1=L-1}=\frac{1}{4}
\Big(\frac{\lambda^2}{\lambda^2+1}\Big)(\delta_{Q0}-\delta_{Q\pi}).
\end{equation}
The explicit formulas for $\Delta(Q,0)_{N=L/2+1}$ and
$\Delta(Q,0)_{N=1}$ show  on one hand the similarity between these
two cases and  on the other that the filled sub-band does have an
effect on the extra particle. Actually as can be seen in
Fig.~\ref{fig2} and \ref{figv},  the negative fringes survive even
for more than one atom in the second band.

Eq.(\ref {eq3}) also shows  that the absence or the presence of an
interference peak at $Qa=\pi$ can be used to distinguish the
fractional and Mott insulating phases: the peak at $Qa=\pi$
disappears when the system is a unit filled insulator but continues
to exist at $aQ=\pi$ when the system is a half-filled insulator. On
the other hand, for both insulating phases the intensity of the
central peak is $\nu(\nu+1)$.

\section{Soft core bosons}
\label{scb}

\begin{figure*}[htbp]
\includegraphics[width=6.5 in]{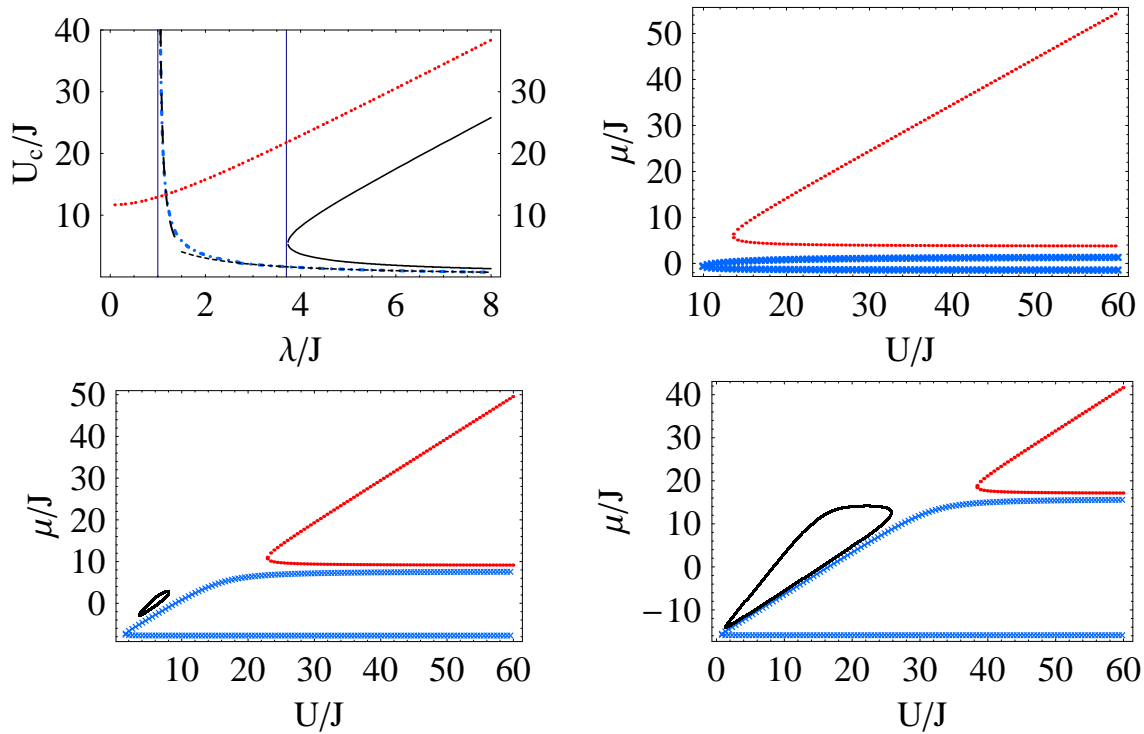}
\leavevmode \caption{(color online) Mean field phase diagram for the
period two superlattice. The  top left panel shows the critical curves, $U_c/J-\lambda_c/J$
for the onset to various transitions to insulating phases. The dash dotted blue line is for the
transition to the band half-filled MI The the long dashed and
short dashed asymptotes corresponds to $2.3/({\lambda}/J-1)$ and
$6.1/({\lambda}/J)$ respectively. The red dotted line corresponds to
the transition to the unit filled MI and the black solid line to the
staggered unit filled insulator. The grid lines are at ${\lambda}=J$
and 3.75. The top right, bottom left and bottom right panels show
the phase diagram as a function of the chemical potential for
systems with ${\lambda}=0.5 J,4 J$ and 8 J} \label{meanph}
\end{figure*}

In this section we relax the hard-core constraint and explore the
interplay between the finite interaction effects  and the competing
periodicity induced by the superlattice potential. The key questions
that we address are: (1) how a finite value of $U$ affects the
various phases observed in the HCB limit, (2) how generic are the
characteristic signals of the phase transitions observed in the HCB
noise correlations as  $U$ becomes finite and (3) what novel
characteristics of the interference pattern emerge as we explore the
$(J,U,\lambda,\nu)$ parameter space.

Before we describe the details of HBTI, we will first use mean field
theory to gain some insight about the phase diagram as
$(U/J,\lambda/J,\nu)$ vary. Although previous studies  based on
numerical Monte carlo simulations have calculated the various phases
in  the soft core regime
\cite{Rigol,Buonsante1,Buonsante2,Buonsante3,Buonsante4}, here we
obtain the phase diagram analytically by using second order
perturbation theory. It is well known that the mean-field
calculations provide a good characterization of the phase diagram
and gives qualitative prediction of  the  different phase
transitions even though the exact transition thresholds may not be
quite correct. We would also like to point out that earlier mean
field studies \cite{Buonsante1,Buonsante2,Buonsante3,Buonsante4} of
the phase diagram were done in a different case, namely when the
hopping parameter is site dependent.

\subsection{Mean-Field Phase Diagram}

To calculate the phase diagram  we use the well know Gutzwiller
approximation \cite{Gutz} that decouples the kinetic energy term of
the Hamiltonian by introducing a superfluid order parameter. The
details of the mean field calculations are discussed in Appendix A.
Mean field theory predicts three insulating phases in addition to
the superfluid phase. The insulating phases can be classified as:
(1) a MI phase with $\nu=1$ where all sites have unit occupancy, (2)
a staggered MI phase with $\nu=1$ characterized by every alternate
site doubly occupied and  (3) a fractional MI state with $\nu=1/2$
where alternate sites are singly occupied. At other filling factors,
the system is always a superfluid.

Fig.~\ref{meanph} summarizes the mean-field calculations. The top
left panel shows  the critical values of $\bar{U}\equiv U/J$ and
$\bar{\lambda}\equiv \lambda/J$ for the onset to the three
insulating phases. They correspond to the tips of the insulating
lobes ( multicritical points) that are shown for selected values of
$\bar{\lambda}$ in the various panels of the figure.

One of the interesting results of the mean-field theory is the
prediction of  a reentrance to the superfluid phase from the
staggered MI phase. In contrast, to the critical curves describing
the threshold to the fractional MI and integer MI phases, which vary
monotonically in the  $(\bar{U},\bar{\lambda})$ space, the curve for
the staggered MI at $\nu=1$ bends (see  left panel in Fig.
\ref{meanph} ) and becomes a double-valued function. As a
consequence, for a fixed $\bar{\lambda}
> \bar{\lambda_c}$, with the threshold value being approximately
$\bar{\lambda}_c=3.75 $,  as we increase the interaction $\bar{U}$,
the system goes from superfluid to staggered MI,  reenters the
superfluid phase and  ends  in  the unit filled  MI phase. Note that
this reentrance to the superfluid phase is absent for
$\bar{\lambda}< \bar{\lambda}_c$, where one only sees a direct
transition from superfluid to unit MI. This behavior is explicitly
shown in the other panels of the figure where the chemical potential
is plotted as a function of the interaction energy, for a few
selected values of $\bar{\lambda}$. In the top right panel
$\bar{\lambda}< \bar{\lambda}_c$  and so only the unit Mott and
fractional Mott phases are present. For
$\bar{\lambda}>\bar{\lambda}_c$,  the staggered phase appears as an
isolated island in the $\mu-U$ plane whose size grows with
$\bar{\lambda}$.

The  numerical Monte-Carlo calculations reported in Ref.
\cite{Rigol} are in agreement with the intermediate superfluid phase
found at mean-field level. However, according with their
calculations the superfluid region is only present for $\bar{U}<12$.
For $\bar{U}>12$ the system is either a unit filled  MI or a
staggered MI. Our mean field calculations, on the other hand,
predict that the intermediate superfluid phase exists for all values
of $\bar{U}$. As we will discuss later, this result is in
qualitative agreement with our numerical calculations performed by
exact diagonalization of the BH Hamiltonian for finite size systems.

The staggered phase describes an interesting manifestation of the
interplay between the onsite interaction energy, $U$, and the energy
modulation introduced by the superlattice potential, $\lambda$. This
phase is absent in the HCB limit and only exists for values of
$\bar{\lambda} \equiv\lambda/J$ beyond a threshold value and for a
finite window of $\bar{U} \equiv U/J$ values. The analytic
calculations predict that the threshold value is
$\bar{\lambda}_c=3.75 $, which is an overestimation of the
corresponding value, $\bar{\lambda}_c\approx 1.75  $, found with the
Monte Carlo simulations \cite{Rigol}.

For the  half filled case, the mean-field critical curve has
$1/\bar{\lambda}$ as an asymptote (see top left panel in Fig.
\ref{meanph}). This scaling  is consistent with Eq. (\ref{sinpa})
(for large $\bar{\lambda}$ the effective tunneling rate between
consecutive odd sites  goes like $J^2/\lambda$). Therefore, for
large $\bar{\lambda}$, a fractional MI state can  exist for rather
small values of  $\bar{U}$. On the other hand Fig. \ref{meanph}
shows that for moderate values of $\bar{\lambda}$ the transition to
an insulating phase requires very large interactions. We would like
to point out that the existence of a fractional Mott phase only for
values of $\bar{\lambda }> 1$ seems to be  an artifact of the
mean-field theory. In fact, in the HCB limit ($U \to \infty$)  the
system is a band insulator for  any infinitesimal value of
$\bar{\lambda}$.

\subsection{Noise Correlations and Phase Diagram}

To establish a correspondence between the various phase transitions
and their signatures in noise-spectroscopy, we resort to exact
diagonalization procedures. We first numerically diagonalize the BH
system to obtain the ground state of the system. This is then used
to obtain two and four point correlations and their Fourier
transform. Here we will describe our results for $8$ wells ($L=8$).

\begin{figure}[tbh]
\begin{center}
\leavevmode {\includegraphics[width=3.5 in]{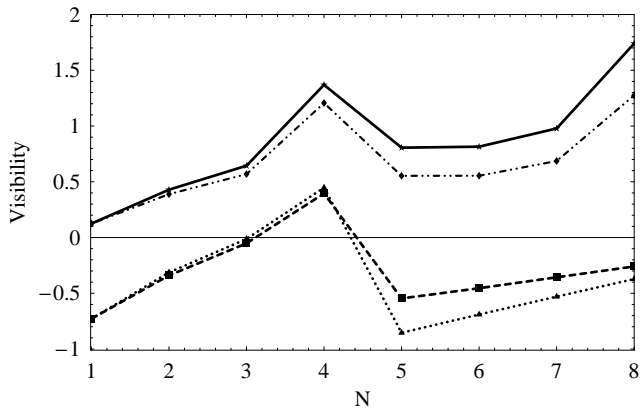}}
\end{center}
\caption{ For fixed $\lambda/J=1$ and $U/J=20$, we plot the
visibility of the central (dotted-dashed line) and superlattice
induced (dashed lines) peaks as the filling factor varies. The solid
and dotted lines are the correspondent  HCB curves}\label{p2dU}
\end{figure}

\begin{figure}[tbh]
\begin{center}
\leavevmode {\includegraphics[width=3.5 in]{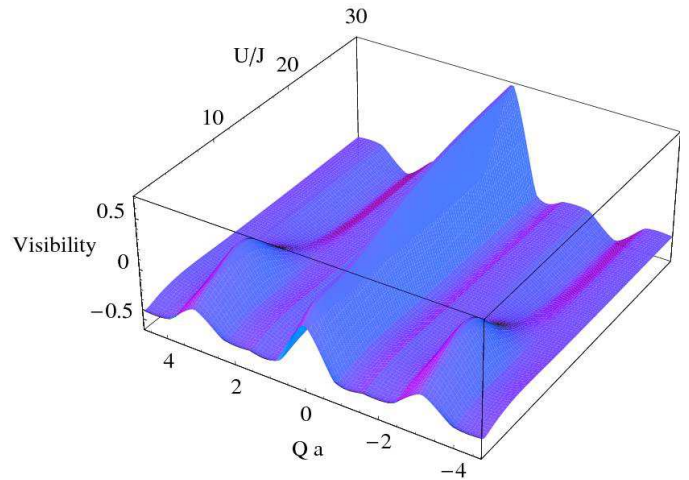}}
\end{center}
\caption{Visibility  plot  (see text for definition)  as a function
of $U/J$ for a system just beyond  half filling ($L=8, N=5$) and for
$\bar{\lambda}=2$.}\label{figpl}
\end{figure}

Fig.~\ref{p2dU} describes the visibility [Eq. (\ref{vis})] of the
second order fringes for various filling factors. The important
point to be noticed here is that the peak to dip transition,  a key
signature of the fractional Mott transition  in the HCB limit, is
preserved for finite $U$. The figure also shows the corresponding
result for the HCB case. A comparison between the $L=80$ (
Fig.~\ref{figv}) and $L=8$ results for HCB illustrates the finite
size effect on the peak to dip transition. As expected, the finite
size of the system broadens the transition and hence reduces the
visibility of the interference peaks.  Fig.~\ref{figpl} further
demonstrates the behavior of the peak to dip effect as the
interaction $U$ is varied. In Fig.\ref{figpl} we plot the normalized
HBTI pattern as a function of $\bar{U}$ for a system just above half
filling. For large but finite $\bar{U}\gtrsim 10$, the superlattice
induced peak in $\Delta(Qa =\pi)$ changes sign and becomes  a dip.
Our numerical analysis confirms that the peak to dip transition is a
generic signature of the strongly interacting regime where bosons
exhibit fermion-type characteristics.

We next show that the second order interference pattern not only
complements the characteristics of the various phase transitions as
observed in the first order Bragg spectroscopy, but it also provides
new definitive signatures of various phases. For clarity, we will
describe our results for $\nu=1$ and  $\nu=1/2$
separately.\\

{\bf (B.I) Unit-Filled Case}
\\

\begin{figure}[tbh]
\begin{center}
\leavevmode {\includegraphics[width=3.7 in]{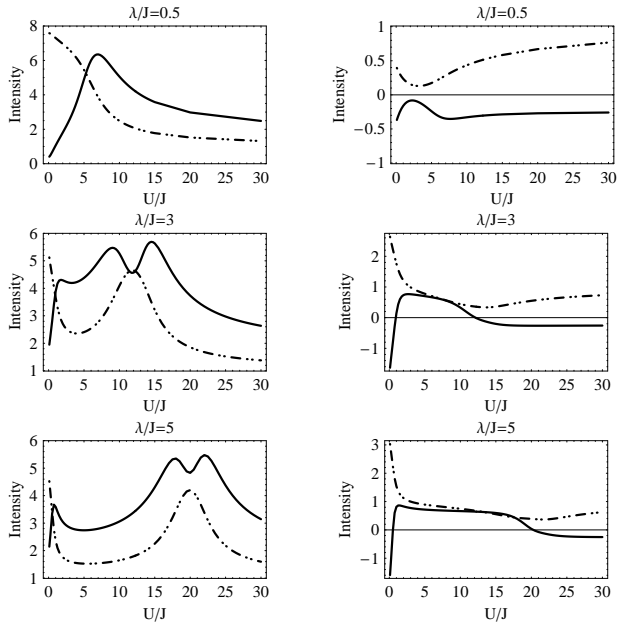}}
\end{center}
\caption{ The figure shows the noise correlations (solid line) and
quasimomentum fridges (dashed line) for various $\bar{\lambda}$
values. The left and right panel respectively describe the
intensities of the  central peak ($Q=0$) and the superlattice
induced peak ($Qa=\pi$).}\label{nomoU}
\end{figure}

Fig.~\ref{nomoU} shows the variation in the central as well as in
the superlattice induced peak for both the first and the second
order interference patterns as the on-site repulsive interaction is
varied. An important aspect of the figure is the appearance of
various local maxima in $\Delta(0)$. The single maximum observed for
small $\bar{\lambda}$  splits as $\bar{\lambda}$ increases. A
comparative study between $N=6$ and $8$ shows that the height at the
various maxima increase with $N$, suggesting that it may be
divergent in the  thermodynamic limit. This observation along with
the fact that the peak for small $\bar{\lambda}$ occurs at a value
of $\bar{U}$ very close to the well known MI transition, suggests
that the maxima may be associated with the onset of the various
superfluid to MI phase transitions.

\begin{figure}[tbh]
\begin{center}
\leavevmode {\includegraphics[width=3.5 in]{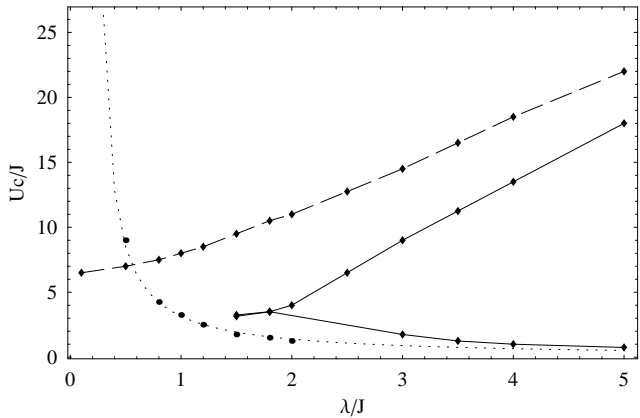}}
\end{center}
\caption{Phase diagram for the period two superlattice calculated by
reading the maxima of the central peak in the $\Delta(0)$ vs.
$\bar{U}$ plot for different  $\bar{\lambda}$ . The dotted, dashed
and  solid lines  show the critical values in the $\bar{U}$ vs
$\bar{\lambda}$ plane for the transition to the  half-filled MI, the
unit filled MI and staggered  filled insulators respectively.
}\label{figph}
\end{figure}

By monitoring the locations of the maxima, we  obtain the phase
diagram in the $(\bar{U},\bar{\lambda})$ plane. Our results are
shown in Fig.~\ref{figph}. The  rather remarkable qualitative
agreement between this phase diagram and that obtained by mean-field
and also by earlier Monte carlo studies, supports the validity of
our conjecture regarding the relationship between a peak in
$\Delta(0)$ and the onset of a phase transition, and suggests that
noise correlations can be used as a practical tool to obtain phase
diagrams of many body quantum systems. It should be noted (Fig.
~\ref{nomoU}) that in contrast to $\Delta(0)$, the  corresponding
zero quasimomentum component  does not provide any sharp signatures
of the different critical points as $U$ is varied.

As shown in the figure for $\bar{\lambda}=3$, for moderate on-site
interaction, the system is  in the superfluid phase. As $\bar{U}$
increases, correlations begin to build up and when $U$ becomes
comparable to $J$, for $\bar{\lambda}>1.5$, the system enters the
staggered Mott insulating phase with two atoms in the low energy
wells and  none in the high energy ones. The onset to this
transition is signaled by a peak in $\Delta(0)$ (as seen in the
figure) and by the development of a positive intensity peak at
$Qa=\pi$. It should be noted that the critical value of
$\bar{\lambda}=1.5$ predicted by this method for the appearance of
the staggered phase is very close to  the value of 1.75 found by the
numerical Monte-Carlo calculations \cite{Rigol}.

The system continues to be in the staggered insulating phase as
$\bar{U}$ increases until the point when the competition between
$\bar{U}$ and the  energy offset induced by the superlattice, $4
\bar{\lambda}$, drives the system back to a superfluid phase. A
simple understanding of this reentrance can be obtained by realizing
that when $\bar{U}$ is of the order of $4 \bar{\lambda}$ the state
$|2020 \dots\rangle$  is degenerated with the states
$|20112011\dots\rangle$, $|10211021\dots\rangle$,
$|21102110\dots\rangle$, and $|11201120\dots\rangle$ which  have
superfluid character as their average densities are $3/2$ and $1/2$
in the low and high energy wells respectively. The reentrance to the
superfluid phase is signaled by the second peak in $\Delta(0)$.
Furthermore, it is accompanied by the disappearance of superlattice
induced peak at $Qa=\pi$. For values  of $\bar{\lambda}$ close to
1.5 the staggered phase exist only for a very narrow range of
$\bar{U}$ values and the first and second order peak can not be
resolved.

 As $\bar{U}$ increases beyond $4\bar{\lambda}$,
it is energetically costly to have two atoms in the same well and
hence the system enters the MI phase where sites are singly
occupied. This transition is also signaled by  the third peak in
$\Delta(0)$. As $\bar{\lambda}$ is increased from
 1.5 up to $\bar{\lambda}\approx 5$, the separation between the last two maxima
 (which determines the range of  $\bar{U}$ values where
the intermediate superfluid phase exists) decreases. Beyond this
value, as $\bar{\lambda}$ is increased further the position of the
two peaks is shifted to larger values of $\bar{U}$.  However,  the
relative separation between the peaks was found to remain constant.
This finding is in disagreement with the results of Monte Carlo
simulations which find  points in $\bar{U}-\bar{\lambda}$ space
where staggered and Mott phases coexist. One may argue that the
absence of  coexistence between the staggered and Mott phase in our
analysis may be due to the finite size of our system. Nevertheless,
the coexistence of such phases seems to imply the existence of a
first order transition. This makes such a coexistence point somewhat
subtle and needs further investigation. Besides this difference, the
phase diagram calculated by monitoring the maxima in $\Delta(0)$ is
in very good qualitative agreement with the Monte carlo as well as
the mean field calculations. The quantitative differences are due to
finite size effects.

We would like to point out that $\Delta(Q=\pi/a)$ also contains
valuable information about the different phases. For example  the
existence of a staggered phase is indicated by the development of a
positive interference peak at $Qa=\pi$. This peak is a second order
effect and it is not present in the quasimomentum distribution. The
quasimomentum distribution does not give definite signatures of the
transition points but it gives information about the phase coherence
of the various phases. As shown in Fig.~\ref {nomoU}, inside the
superfluid phases the $n(0)$ vs $\bar{U}$ develops  a maximum which
disappears in the insulating phases where the curve tends to become
flat.
\\

{\bf (B.II) Half filled Case}
\\

\begin{figure}[tbh]
\begin{center}
\leavevmode {\includegraphics[width=3.7 in]{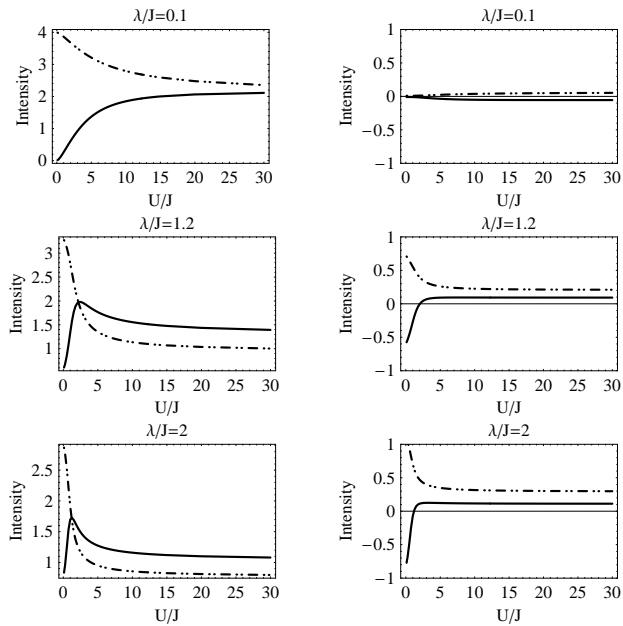}}
\end{center}
\caption{ The figure shows the noise correlations (solid line) and
quasimomentum distribution (dashed line) for $\nu=1/2$ and  various
$\bar{\lambda}$ values. The left and right panel respectively
describe the intensities of the central peak $Q=0$ and the
superlattice induced peak ($Qa=\pi$).}\label{nomoUH}
\end{figure}

We next discuss the $\nu=1/2$ case, where one sees a transition from
superfluid to fractional Mott phase. The formation of this phase is
also signaled by the development of a sharp peak in the intensity of
the central noise correlation peak as $ \bar{U}$ is varied across
the transition.  This is shown in Fig. \ref{nomoUH}. In our finite
size system,  the peak was found to exist only for values of
$\bar{\lambda}\gtrsim 0.205$. However, we expect this value to
decrease as the size of the system increases. Fig. \ref{figph} shows
the position of the peak as a function of $\bar{\lambda}$. For large
$\bar{\lambda}$ the critical $\bar{U}$  value decreases as $1/
\bar{\lambda}$ in consistence with the mean field results.

The formation of the half filled band insulator is also indicated by
the superlattice induced peak at $\Delta(\pi/a)$. It should be noted
that $\Delta(\pi/a)$ has negative intensity for the non-interacting
system. Precisely  at $U=0$ $\Delta(\pi/a)=-N/4
\lambda^2/(\lambda^2+J^2)$. Its amplitude increases as $\bar{U}$ is
increased and becomes positive for the values of $\bar{\lambda}$
that allow the formation of a band-insulator. For
$\bar{U}>\bar{U}_c^{\nu=1/2}$, with $\bar{U}_c^{\nu=1/2}$ the Mott
band insulator critical point, the peak intensity almost reaches its
hard-core value.

\section{Trapped System}
\label{trap}

In this section we study the experimentally relevant case of a
superlattice in the presence of an additional parabolic potential.
The two key results of this study are  following: First, to find
that in contrast to the homogeneous system where the MI phases only
exist at specific critical fillings, the trapping potential
stabilizes the MI phases and allows for their existence over a large
range of filling factors. Second, to show that  noise-spectroscopy
provides a detailed information of the formation of the insulating
domains and the shell structure that arises in presence of a
parabolic confinement.

Understanding the effects of the trap on period-$2$ superlattice
follows along the lines of the earlier studies of the trapped system
in the absence of any additional lattice modulation. Therefore, we
first  briefly review the key aspects of the $\lambda=0$ case which
has been studied in great detail \cite{Rigol1,Quintanilla,Ott,Rey}.
The combined lattice plus harmonic confinement system possesses two
distinct classes of eigenstates: low energy states that extend
symmetrically around the trap center and high energy states that are
localized on the sides of the potential. The origin of these two
distinct classes is related to the two energy scales in the system,
namely the tunneling ($J$) and the trapping energy ($\Omega$). Modes
with excitation energy below $4J$, which correspond to the band
width of the translationally invariant system, are extended and can
be thought of as harmonic oscillator like modes with effective
frequency $\omega^*=\sqrt{4 J\Omega}$ and effective mass
$m^*=\hbar/(2 J a^2)$. Modes with excitation energies above $4J$ are
close to position eigenstates since for these states the kinetic
energy required for an atom to hop from one site to the next becomes
insufficient to overcome the potential energy cost. The high energy
eigenstates are almost two-fold degenerate with energy spacing
mostly determined by $\Omega$. The localization of these modes can
be understood by means of a semiclassical analysis
\cite{Quintanilla}. Within WKB scheme, the localization of the
higher energy modes can be linked to the appearance of new turning
points related to umklapp processes. In contrast to the turning
points of the classical harmonic oscillator that appear at zero
quasimomentum, the Bragg turning points emerge when the
quasimomentum reaches the end of the Brillouin zone and can
therefore be associated with Bragg scattering induced by the
lattice.

\begin{figure}[htbp]
\includegraphics[width =1\linewidth]{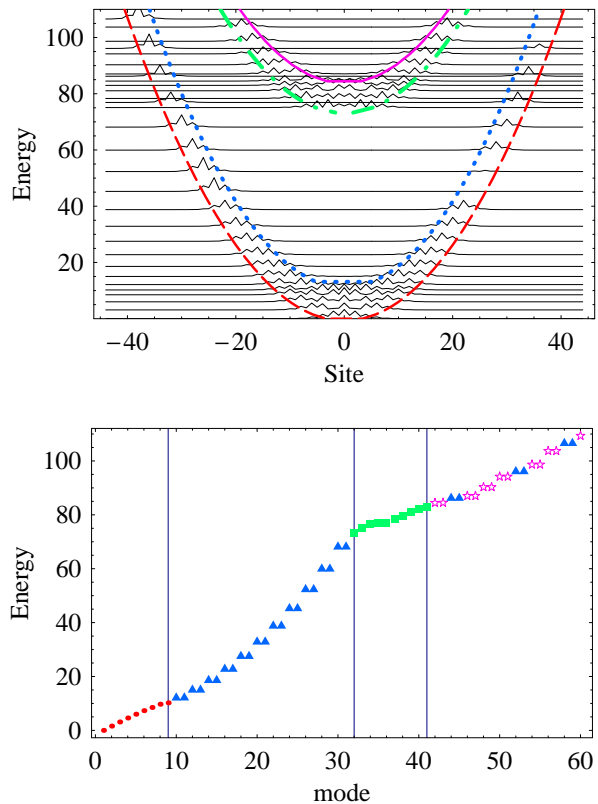}
\leavevmode \caption{(color online) The top panel shows the
single-particle eigenstates of a system with $\lambda/J =1.1$ and
$\Omega/J=0.005$. The height of each mode is  proportional to its
energy in units of $\sqrt{\Omega J}$. The parabola correspond to the
classical and Bragg turning points that appear in the first and
second band respectively (see text). In the lower panel we show the
spectrum of the modes in  units of $\sqrt{\Omega J}$. The dots,
 triangles,  boxes and stars  indicate the low energy modes in the
first band, the high energy localized modes in the first band, the
low energy modes in the second band and  high energy localized modes
in the  second respectively. Note the degeneracy that appears
between the localized high energy modes of  the two bands.
 The grid lines are at $N_1^{LE}=9$, $N_1=32$ and
$N_2^{LE}=42$. } \label{figtm}
\end{figure}

The period-$2$ superlattice splits a band into two subbands and hence
the nature of modes can be now classified by the energy scales associated
with these two bands and the band gap which is equal to $4\lambda$.

In the top panel of Fig.\ref{figtm} we plot the various wave
functions. Each eigenstate has been  offset along the $y$ axis by
its energy in units of $\sqrt{J \Omega}$.  In the lower panel we
also show the energy spectrum. As shown in the figure, we now see
two sets of extended and also two sets of localized states. Again,
the localized states are related to umklapp processes and to
emphasize this idea,  the classical and the Bragg turning points of
both bands are explicitly displayed. In Appendix B, we provide
additional details of the trapped superlattice model.

\begin{figure}[htbp]
\includegraphics[width =0.8\linewidth]{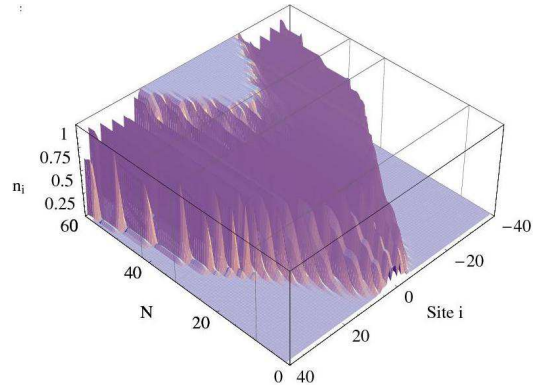}
\leavevmode \caption{(color online) Density distribution for a
system with $\bar{\lambda} =1.1$ and $\Omega/J=0.005$ as a function
of the total number of atoms.The grid lines are at $N_1^{LE}=9$,
$N_1=32$ and $N_2^{LE}=42$. } \label{figden}
\end{figure}

\begin{figure}[htbp]
\includegraphics[width =0.8\linewidth]{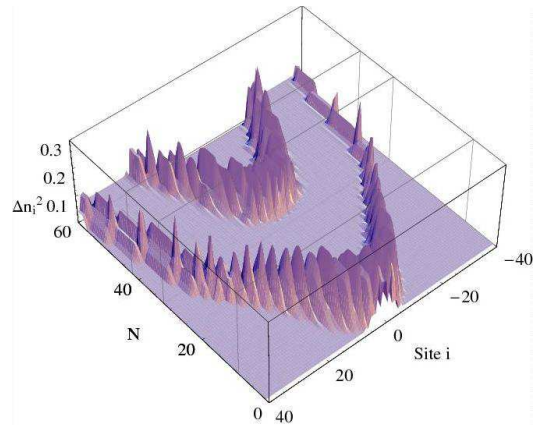}
\leavevmode \caption{(color online) Number fluctuations in the
presence of trap for a system with $\bar{\lambda} =1.1$ and
$\Omega/J=0.005$ as a function of the total number of atoms. The
grid lines are at $N_1^{LE}=9$, $N_1=32$ and $N_2^{LE}=42$. }
\label{figflu}
\end{figure}

Figs.~\ref{figden} and \ref{figflu} describe the density and number
fluctuations that reflect the band structure of the trapped
superlattice system discussed above.  To simplify the description,
we introduce three different quantities: $N_1^{LE}$, $N_1$ and
$N_2^{LE}$ which respectively denote the number of low energy
extended states in the first band, the number of states below the
first mode of  the second band  and the total number of modes below
the first localized mode of  the second band. In Fig.~\ref{figtm},
these three number are explicitly indicated with grid lines.

For filling factors below $N_1^{LE}$  atoms tend to spread over the
central sites populating mostly the low energy wells. The extended
character of the modes induce  large number fluctuations. For
filling factors between $N_1^{LE}<N \leq N_1$, the localized modes
in the first band become occupied. Because these modes are localized
at the edges of the cloud, as the number of atoms is increased the
occupation of the central site remains constant and instead sites
farther away from the trap center become populated. Thus the
presence of  single-particle localized modes in the Fermi sea leads
to the formation of fractionally filled insulating domains at the
trap center in the many-body system. Alternatively,  as soon as
localized modes are populated, number fluctuations become only
relevant at the edges. The insulator character of the atoms at these
central sites and the reduced number fluctuations  can be seen in
Fig.\ref{figflu}.

For $N>N_1$ the trapping energy cost of placing an atom at the trap
edge is higher than the energy needed to place the atom at the
center. For $1<N<N_2^{LE}$,  atoms occupy the extended modes of the
second band and tend to spread over the  different  empty sites at
the trap center. When the number of atoms $N=N_2^{LE}$, the first
localized mode of the band is populated and  the site at the trap
center acquires filling factor one. As $N$ is further increased, the
width of the unit-filled central core grows. For $N>N_2^{LE}$ the
density profile alternates from unit filled MI, superfluid,
fractional
 MI  and  superfluid as one moves from the center
towards the edge of the cloud. This shell structure in the HCB limit
resembles in many aspects the shell structure observed in the
$\lambda=0$ case at moderate values of $U$. In the former case, the
gap is induced by the external modulation and the shells consist of
fractional and unit filled insulator domains surrounded by
superfluid regions. In the later case, the gap is induced by the onsite
interaction energy  $U$  and the shells consist of integer filled MI
domains surrounded by superfluid regions.

\begin{figure}[htbp]
\includegraphics[width =0.9\linewidth]{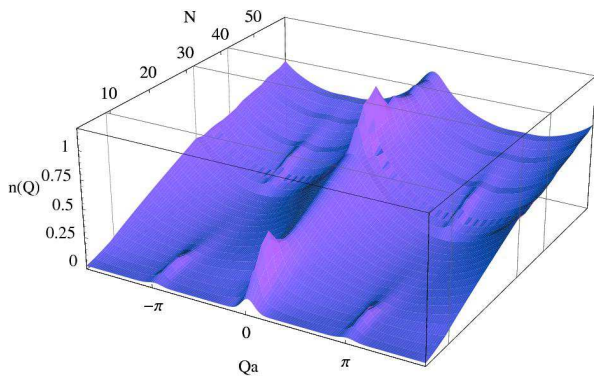}
\leavevmode \caption{(color online) Quasimomentum distribution for a
system with $\lambda/J =1.1$ and $\Omega/J=0.005$ as a function of
the filling factor. The grid lines are at $N_1^{LE}=9$, $N_1=32$ and
$N_2^{LE}=42$. } \label{figmom}
\end{figure}

We now describe the first and second order interference patterns in
the trapped system. The quasimomentum distribution is  shown in
Fig.~\ref{figmom}. Consistent with the behavior observed in the
density and number fluctuations, for filling factors $N<N_1^{LE}$
and  $N_1< N<N_2^{LE}$ the quasimomentum distribution is sharply
peaked at $Q=0$ and $Qa=\pi$, signaling the superfluid character of
the system. The insulating phases that appear for filling factor
$N_1^{LE}<N<N_1$ and $N>N_2^{LE}$ are signaled in the quasimomentum
distribution  by the drop of the peak intensities and the flattening
of the quasimomentum distribution profile.

\begin{figure}[htbp]
\includegraphics[width =1\linewidth]{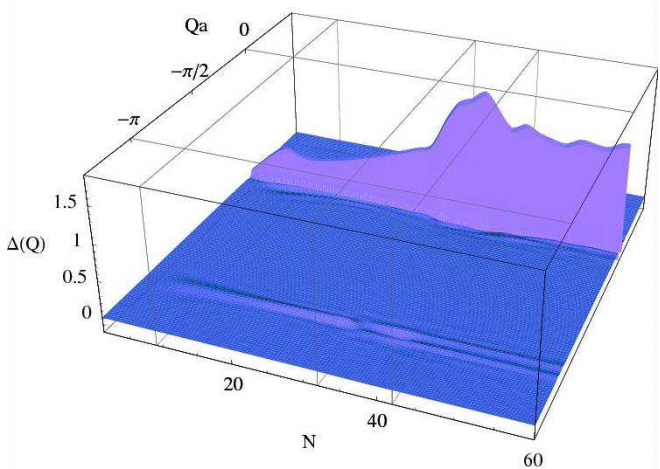}
\leavevmode \caption{(color online) Noise interference pattern as a
function of the filling factor or a system with $\lambda/J =1.1$ and
$\Omega/J=0.005$. The grid lines are at $N_1^{LE}=9$, $N_1=32$ and
$N_2^{LE}=42$. } \label{figno}
\end{figure}

Fig. \ref{figno} shows the  HBTI  pattern for a different total
number of atoms. For $N<N_1^{LE}$, only extended modes are occupied
and the intensity of the central peak grows monotonically with $N$.
Similarly the peak at $Qa=\pi$ increases from the negative value it
takes for $N=1$ and becomes a peak  with  a positive intensity. As
the localized modes enter the Fermi sea ($N> N_1^{LE}$) the rate of
growth of the central peak slows down and this change in the slope
generates a maximum in the interference pattern. Note that in
contrast to the strong reduction observed in the   $\Omega=0$  HCB
case, in the presence of the trap only a decrease in the rate of
growth is observed. The reason of this behavior is the fact that in
the trapped case there is always a superfluid component at the edges
of the atomic  cloud. For $N_1^{LE}> N \geq N_1$ the interference
peak at $qa=\pi$ is clearly visible, reflecting the staggered
character of the phase.

At $N=N_1+1$, we see a rather sharp peak to dip transition at
$Qa=\pi$. Similar to the homogeneous system, the transition signals
the beginning of  the population of an empty band. The dip becomes a
peak when the number of atoms increases beyond a certain value.
For $N_1 < N \leq N_2^{LE}$, one also observes an
increase in the growth rate of the central peak with N, consistent
with the superfluid properties of the system at these fillings.
Finally for $ N > N_2^{LE}$, a unit filled insulating domain at
appears at the trap center. This leads to a decrease in the amplitude of the
central peak until it  reaches a constant value. The peak at
$Qa=\pi$ disappears as the unit filled core at the trap center
grows. In the regime  $ N > N_2^{LE}$, Fig. \ref{figtm} shows some
additional small modulations of the central peak amplitude at
certain fillings. The oscillations take place when a high energy
mode of the first band, degenerated with another high energy mode of
the second band, enters the Fermi sea.

\begin{figure}[htbp]
\includegraphics[width =0.7\linewidth,height=0.7\linewidth]{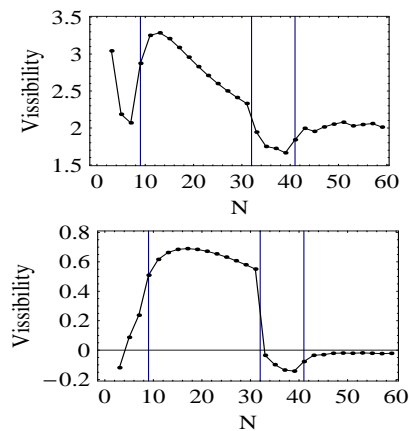}
\leavevmode \caption{Visibility of the second-order interference
peaks at $Q=0$ (top panel) and $Qa=\pi$ (bottom panel)
 as the total number of trapped atoms  is varied.  The grid lines are at $N_1^{LE}=9$, $N_1=32$ and
$N_2^{LE}=42$.
  }\label{figptodip}
\end{figure}

In analogy with the homogeneous case, the transition to insulating
state can best be illustrated in the normalized intensity of the
noise correlations, namely the visibility. In Fig.~\ref{figptodip},
we plot the visibility of the central and the superlattice peaks. As
mentioned above, the normalized pattern is perhaps the best
experimental observable as the normalization procedure filters some
of the technical noise introduced in the measurement procedure. The
normalization procedure maps the central peak maxima that appear
before the formation of the insulator domains to minima.
Furthermore, when normalized, the amplitude of the superlattice
induced peak becomes comparable in intensity to that of the central
peak.

\section{Conclusions}
\label{conc}

In this paper, we have studied shot noise correlations for
interacting bosons in the presence of an additional lattice.
Although we have focused on a period-$2$ superlattice,  this simple
case  exhibits all the key aspects of the phase diagram of more
general superlattice potentials.

One of the central results of this paper is that noise correlations
may provide a practical tool to construct the  phase diagram of many
body quantum systems exhibiting transitions from superfluid to
fractional and integer Mott phases which also include the staggered
insulating phase. Our study suggests that the intensity of the
second order Bragg peaks provides a new order parameter to
characterize these phase transitions. Although our calculations in
the soft core regime were done for systems with a reduced number of
atoms and wells and further studies for larger systems are needed to
confirm our conjecture, we believe that this result is important in
the theory of quantum phase transitions. Furthermore, our study of
the trapped system demonstrates the possible realization of these
phases in the laboratory.

Shot noise spectroscopy has already proven to be a useful
experimental tool to identify Mott insulating states. We consider
our analysis of the shot noise and its possible relevance in
identifying various phase transitions in interacting bosons in
superlattices is pertinent for  atom optics experiments. Up to date
all the experiments in super-lattice has been restricted to the
Bose-Einstein-condensate regime. However, ongoing experimental
efforts are trying  to reach the strongly correlated regime in
superlattices. The motivation of these efforts is not only to gain
understanding of the many-body physics
 but also because  atoms selectively loaded in super-lattices
can have wider separation and stronger confinement,  useful
properties for the implementations of lattice based quantum
computing proposals. We hope that our study will stimulate further
experimental probing of one dimensional bosonic systems loaded in
superlattices via HBTI.

\section{Appendix A}

To study the phase diagram for the homogeneous system we use the
well know decoupling approximation \cite{Gutz} and substitute:
\begin{eqnarray}
\hat{a}_i^{\dagger}\hat{a}_{j}=\psi_i \hat{a}_{j}+\psi_j
\hat{a}_{i}^{\dagger}-\psi_i\psi_j
\end{eqnarray}into the BH Hamiltonian. Here $\psi_j=\langle\hat{a}_{j}\rangle \approx \sqrt{n_j}$  is the superfluid order
parameter and $n_j$  is the expectation value of the number  of
particles on site $i$. We will assume  the order parameter  to be
real. This substitution leads to :

\begin{eqnarray}
\hat{H}^{ef}&=&-J\sum_j (\psi_{j+1}+\psi_{j-1})
(\hat{a}_{j}^{\dagger}+\hat{a}_j-\psi_{j})\\&&\notag+ \sum_{j}(2
\lambda \cos[\pi j] -\mu) \hat{n}_j + \frac{U}{2}\sum_{j}
\hat{n}_j(\hat{n}_j-1), \label{Hef}
\end{eqnarray} where we have also introduced the chemical potential
$\mu$. Assuming  periodic boundary conditions, one can reduced the
problem  to  a two mode system due the fact that  the Hamiltonian is
exactly the same for all even and all odd sites. We define
$\psi_2=\psi_{2i}$, $\psi_1=\psi_{2i+1}$, $\hat{a}_2=\hat{a}_{2i}$
and $\hat{a}_1=\hat{a}_{2i+1}$, so

\begin{eqnarray}
\hat{H}^{ef}&=&\frac{N}{2}(\hat{H}^{ef}_1+\hat{H}^{ef}_1),\\
\hat{H}^{ef}_1&=& -2J \psi_2
(\hat{a}_{1}^{\dagger}+\hat{a}_1-\psi_1)\\&&\notag- (2 \lambda
 +\mu) \hat{n}_1 + \frac{U}{2}
\hat{n}_1(\hat{n}_1-1),\\
\hat{H}^{ef}_2&=& -2J \psi_1
(\hat{a}_{2}^{\dagger}+\hat{a}_2-\psi_2)\\&&\notag+ (2 \lambda
 -\mu) \hat{n}_2 + \frac{U}{2}
\hat{n}_2(\hat{n}_2-1). \label{Hef}
\end{eqnarray}
The above Hamiltonian can be diagonalized using a Fock state basis
truncated  below a certain occupation value. However, second order
perturbation theory using $\psi$ as an expansion parameter can
provide a good analytic approximation of  the phase diagram. At zero
order in $\psi$ the Hamiltonian is diagonal in the Fock basis. The
integer occupation numbers $n_1,n_2$ that minimized the energy are
given by the conditions
\begin{eqnarray}
n_1-1<\frac{\mu+2\lambda}{U}<n_1,\\
n_2-1<\frac{\mu-2\lambda}{U}<n_2.
\end{eqnarray}
We consider three different cases:

\begin{itemize}
  \item $-2\lambda<\mu<\min(U-2\lambda,2\lambda)$
\end{itemize}

 In this case the
ground state of the system is the Fock State $|010101\dots\rangle$
and the unperturbed energy is $E_g^{(0)}=-2\lambda -\mu$. The first
order correction to the energy vanishes and  to second order one
gets:
\begin{eqnarray}
E_g^{(2)}=4J\left(\frac{J\psi_1^2}{\mu-2\lambda}- \frac{J
\psi_2^2}{\mu+2\lambda}+\frac{2J\psi_2^2}{\mu+2\lambda-U}+\psi_1\psi_2
\right).
\end{eqnarray} Minimizing the energy respect to $\psi_{1,2}$ yields the following equation:
\begin{eqnarray}
\left(
  \begin{array}{cc}
    1 & \frac{2J}{\mu-2\lambda} \\
     \frac{2J( \mu+2\lambda+U)}{(\mu+2\lambda-U)(\mu+2\lambda)} & 1 \\
  \end{array}
\right)\left(\begin{array}{c}
               \psi_2 \\
               \psi_1
             \end{array}
\right)=\left(\begin{array}{c}
               0 \\
               0
             \end{array}
\right).\label{order}
\end{eqnarray} Eq. (\ref{order}) has nontrivial solution only when the determinant of the matrix vanishes.
Therefore the surface  where the determinant vanishes determines the
insulating phase:

\begin{eqnarray}
4J^2( \bar{\mu}+2\bar{\lambda}+\bar{U})=(
\bar{\mu}+2\bar{\lambda}-\bar{U})(\bar{\mu}^2-4\bar{\lambda}^2),
\label{mufra}
\end{eqnarray}
where  the bar denotes the dimensionless variables $\bar{U}=U/J$,
$\bar{\lambda}=\lambda/J$ and $\bar{\mu}=\mu/J$. In  Fig.
\ref{meanph} we show, with a crossed blue line, the solutions of Eq.
(\ref{mufra}) in the $\bar{\mu}$ vs. $\bar{U}$ plane,  for
$\bar{\lambda}=0.5$, $4$ and $8$  (top right, bottom left panel and
bottom right panels respectively). The region inside the blue loops
corresponds to the $1/2$ filled insulator.  In the top left panel we
also  show  with a blue dash-dotted line  the critical value of
$\bar{U}_c^{\nu=1/2}$  as a function of $\bar{\lambda}$.
$\bar{U}_c^{\nu=1/2}$  corresponds to the smallest $\bar{U}$ in the
lobe, below which the system is always a superfluid. Exactly  at
$\bar{U}^{\nu=1/2}_c$ the upper and lower branches of the chemical
potential merge and the energy gap closes up.  It is important to
point out that in general mean field calculations do not accurately
predict critical values but, in general, they capture very well the
physics of the phase transition.

\begin{itemize}
  \item $2\lambda<\mu<U-2\lambda$
\end{itemize}
In this case the ground state of the system is the unit filled state
with exactly one atom per site. The zero order ground state energy
is $E_g^{(0)}=-2\mu$. For this case we set  $\psi_1=\psi_2 =\psi$.
To second order the ground state energy is given by:

\begin{eqnarray}
E_g^{(2)}&=& \left(\frac{8 J^2\psi^2}{\mu-2\lambda-U}+\frac{8
J^2\psi^2}{\mu+2\lambda-U}\right)\notag \\&&-\left( \frac{4J^2
\psi^2}{\mu+2\lambda}+\frac{4J^2\psi^2}{\mu-2\lambda}\right)+
4J\psi^2.
\end{eqnarray}Minimizing with respect to $\psi$  one gets an algebraic
equation that determines the boundary between the superfluid and
insulating phases. The solution is displayed in  Fig.\ref{meanph}
for $\bar{\lambda}=0.5, 4$ and $8$.  The critical value of
$\bar{U}^{\nu=1}_c$ is also shown in the top left panel with a
dotted red line. Note  the critical value increases with
$\bar{\lambda}$. This increase is in agreement with the idea that
for weak interactions disorder helps to delocalize the atoms.

\begin{itemize}
  \item $U-2\lambda<\mu<2 \min(U-\lambda,\lambda)$
\end{itemize}
In this case the on-site repulsion is not large enough to avoid
double occupancies and the ground state of the system is the state
 with two atoms in the low energy sites  and zero in the others. This
situation can not be described in the hard core regime and it is
only present at moderate values of $U$. To  zero order in J, the
ground state energy is $E_g^{(0)}=-2\mu-4\lambda+U$.  To second
order it is given by

\begin{eqnarray}
E_g^{(2)}&=&\frac{4J^2\psi_1^2}{\mu-2\lambda-U}-\frac{4J^2\psi_1^2}{\mu+2\lambda}+
\frac{4J^2
\psi_2^2}{2\lambda-\mu}+\frac{8J^2\psi_2^2}{\mu+2\lambda-U}\notag\\&&+4J\psi_1\psi_2
\end{eqnarray}By first minimizing with  respect to $\psi_{1,2}$, and finding   the
solutions for which the determinant vanishes one obtains the
 boundary for this phase. The solution is displayed in Fig.\ref{meanph}
for $\bar{\lambda}=0.5,4$ and $8$ with a black line. At mean field
level the minimum value of $\bar{\lambda}$ required for the
existence of this insulating phase is $\bar{\lambda}=3.75$. For
values of $\bar{\lambda}> 3.75$ the  unit filled system is a
superfluid for  $\bar{U}<\bar{U}^{\nu=1}_{c1}$, a staggered
insulator for $\bar{U}^{\nu=1}_{c1}<\bar{U}< \bar{U}^{\nu=1}_{c2}$
and a unit filled Mott insulator  for $ \bar{U}>
\bar{U}_{c}^{\nu=1}$.  In the top right panel, the  lower and upper
branches of the solid black curve correspond to
$\bar{U}_{c1}^{\nu=1}$ and $\bar{U}_{c2}^{\nu=1}$ respectively.

\section {Appendix B}

The super-lattice potential splits the main band into two different
sub-bands. In this "multi"-band picture, it is simple to understand
the modification introduced by the trap.   For excitation energies
below the band width of the first sub-band: $E^{(n)}-E^{(0)}<
E_{with}^{(1)}\equiv 2\sqrt{J^2+\lambda^2}-2\lambda$, the modes are
delocalized states that spread every other site symmetrically around
the potential minimum.  In the semiclassical picture these modes
only see the classical turning point at $x_{cla}^{(1)}= \pm a
\sqrt{(E^{(n)}+ 2\sqrt{\lambda^2+J^2})/\Omega}$. For higher
energies, $E^{(n)}-E^{(0)}\geq E_{with}^{(1)}$, the eigenstates
become localized on both sites of the potential. These modes see
besides the classical turning point, the Bragg turning point of the
first band $x_{B}^{(1)}= \pm a \sqrt{(E^{(n)}+ 2 \lambda)/\Omega}$.

At  excitation energies higher than $4\lambda$ (the energy gap
between the two subbands), the potential energy cost of localizing a
state at the edge is larger than the energy required to populate the
high energy wells at the trap center. As a consequence, states with
quantum number $n>\sqrt{4 \lambda/\Omega}$ appear centered again
around the trap minima.  These second group of extended states see
the classical turning point at $x_{cla}^{(2)}= \pm a
\sqrt{(E^{(n)}-4\lambda +2\sqrt{\lambda^2-J^2})/\Omega}$. Finally,
for excitation energies larger than the band width of the second
band $E^{(n)}-E^{(0)}\geq E_{with}^{(2)}=E_{with}^{(1)}$, the modes
feel   the Bragg turning point of  the band $x_{B}^{(2)}= \pm a
\sqrt{(E^{(n)}- 2 \lambda)/\Omega}$ and become again localized at
the edge. Sometimes a high energy mode in the second band  becomes
degenerated with some of the very high energy modes of the first
band. This effect can be seen in Fig. \ref{figtm}.

\noindent\textbf{Acknowledgments} We would like to thank  J.V.
Porto, I.B. Spielman and S. Sachdev for their suggestions, comments
and useful input. This work is supported in part by the Advanced
Research and Development Activity (ARDA) contract and the U.S.
National Science Foundation through a grant PHY-0100767. A.M.R.
acknowledges support by a grant from the Institute of Theoretical,
Atomic, Molecular and Optical Physics at Harvard University and
Smithsonian Astrophysical observatory.

\end{document}